\documentclass{jfm}
\pdfoutput=1
\usepackage{epsfig}

\usepackage{amsfonts}
\usepackage{amsmath,amssymb}
\usepackage{cancel,verbatim}
\usepackage{graphics,float}
\usepackage{color}
\usepackage{bm}

\newcommand{\beq}{\begin{equation}}  
\newcommand{\eeq}{\end{equation}}  
\newcommand{\bea}{\begin{eqnarray}}  
\newcommand{\eea}{\end{eqnarray}}

\newcommand{\eps}{\varepsilon}
\newcommand{\epsin}{\varepsilon_{\rm IN}}
\renewcommand{\vec}[1]{\bm{#1}}
\renewcommand{\Rey}{{\rm Re}}
\newcommand{\fvec}[1]{\hat{\bm{#1}}}

\shorttitle{Nonuniversal condensate formation in 2d turbulence}
\shortauthor{M. Linkmann, M. Hohmann, B. Eckhardt}

\title{Nonuniversal transition to condensate formation in two-dimensional turbulence}
\author{Moritz Linkmann
\corresp{\email{moritz.linkmann@physik.uni-marburg.de}},
Manuel Hohmann,
Bruno Eckhardt\corresp{Deceased on the $7^{\rm th}$ of August 2019.}}
\affiliation{
Fachbereich Physik, Philipps-University of Marburg,  D-35032 Marburg, Germany \\
}

\begin{document}
\maketitle 

\begin{abstract}
The occurrence of system-scale coherent structures, so-called condensates, is a
well-known phenomenon in two-dimensional turbulence.  Here, the
transition to condensate formation is investigated as a function of the
magnitude of the force and for different types of forcing.  Random
forces with constant mean energy input lead to a supercritical
transition, while forcing through a small-scale linear instability
results in a subcritical transition with bistability and hysteresis.
That is, the transition to condensate formation in two-dimensional
turbulence is nonuniversal.
For the supercritical case we quantify the effect of large-scale friction on the
value of the critical exponent and the location of the critical point. 
\end{abstract}
\maketitle

\section{Introduction}
\label{sec:intro}

Two-dimensional (2d) and quasi-2d flows occur at macro- and mesoscale in a
variety of physical systems. Examples include stratified layers in Earth's
atmosphere and the ocean \citep{Vallis06}, soap films and more recently also
dense bacterial suspensions, where the collective motion of microswimmers
induces patterns of mesoscale vortices \citep{Dombrowski04,Dunkel13PRL,Gachelin14}. A characteristic
feature of 2d turbulence is the occurrence of an inverse energy cascade
\citep{Kraichnan67,Boffetta14ARFM}, whereby kinetic energy is
transferred from small to large scales. In confined systems, this self-organisation 
can result in the formation of large-scale coherent structures 
\citep{Kraichnan67,Hossain83,Sommeria86,Smith93}, so-called condensates
\citep{Smith93}, which emerge in different forms depending on geometry and
boundary conditions, e.g. as vortex dipoles or jets \citep{Bouchet09,Frishman2017}.

The inverse energy cascade in 2d turbulence is connected with an additional inviscid 
conservation law, that of enstrophy. However, inverse cascades and thus condensates 
are not specific to 2d phenomena. They occur whenever fluctuations in one spatial coordinate are suppressed, 
as is the case in thin fluid layers \citep{Xia09,Celani10,Musacchio17}, or, for instance, in presence of 
rapid rotation \citep{Rubio14,Deusebio14,Gallet15b}, stratification \citep{Sozza15} or both \citep{Marino13}, 
and in presence of a strong uniform magnetic field \citep{Gallet15a} for weakly conducting flows.
Another, fully three-dimensional (3d), mechanism that leads to inverse energy transfer is 
breaking of mirror-symmetry \citep{Waleffe93,Biferale12}. 
In magnetohydrodynamic turbulence the latter can result in the formation of magnetic 
condensates through large-scale dynamo action or the inverse cascade 
of magnetic helicity \citep{Frisch75,Pouquet76}.
%

There is thus a variety of systems that potentially allow mean energy transfer  
from small to large scales, which leads to a variety of possible transitions to 
spectral self-organisation and large-scale pattern formation, the nature of which 
depends on the details of the system, and smooth, supercritical and subcritical
transitions between non-equilibrium statistically steady states have been observed. 
In 3d rotating domains for example, the nature of the transition 
between forward and inverse energy transfer with respect to the rotation rate 
depends on the mechanism by which the condensate saturates \citep{Seshasayanan18}. The two saturation scenarios are:  
(i) saturation by viscous effects as in 2d, where the condensate becomes sufficiently
energetic for the upscale flux to be balanced by viscous dissipation, or 
(ii) saturation by local cancellation of the rotation rate by the counter-rotating
vortex that forms part of the condensate \citep{Alexakis15}. In case (i) the transition
is supercritical \citep{Seshasayanan18}, and in (ii) it is subcritical 
\citep{Alexakis15,Yokoyama17,Seshasayanan18}, showing bistability and
hysteresis \citep{Yokoyama17}. Similar results have been obtained if the magnitude of the forcing is used as a control parameter 
at a fixed value of the rotation rate \citep{Yokoyama17}, with random and static forcing 
both resulting in a subcritical transition. The latter was interpreted as evidence in support
of universality. Hysteretic transitions and bistable scenarios also occur in thin layers 
as a function of the layer thickness \citep{vanKan19}. Subcriticality in the transition to condensate 
formation in rapidly rotating Rayleigh-B\'enard convection 
has been connected with non-local energy transfer from the driven scales into the condensate  
due to persistent phase correlations \citep{Favier19}.

In summary, transitions in cascade directions from direct to inverse and vice versa have received 
considerable attention in recent years, \citet{Alexakis18} provide a comprehensive overview thereof. In contrast, transitions to condensate formation 
in purely 2d turbulence have been studied only in the context of active matter, where 
spatiotemporal chaos and classical 2d turbulence with a condensate are connected 
by a subcritical transition \citep{Linkmann19a,Linkmann19b}. 
Here, we extend this work and focus on the transition to condensate formation in 
two-dimensional turbulence as a function of the intensity the driving and in presence of 
large-scale friction. Conceptually, the 2d geometry
differs substantially from thin layers or rapidly rotating 3d domains, 
as the energy transfer is now purely inverse. 
That is, the transition investigated here does not occur between two non-equilibrium 
statistically steady states with different multiscale dynamics. Instead, in 2d one state 
has multiscale dynamics and the other is a spatiotemporally chaotic state concentrated at 
the driven scales. 
Hence the transition in 2d is towards and away from multiscale dynamics, not between
different types of such. 
By means of direct numerical simulations we show that the nature of the transition 
depends on the type of driving: It is supercritical for random forcing and 
subcritical if the driving is given by a small-scale linear instability.   
In the former case we also explore the effect of large-scale friction on the 
location of the critical point and the value of the critical exponent.


\section{Numerical details}
\label{sec:numerics}
We consider the 2d Navier-Stokes equations for incompressible flow in a square domain $V$ embedded in the $xy$-plane
with periodic boundary conditions. 
In this case, the Navier-Stokes equations can be written in vorticity form
\beq
\label{eq:momentum}
\partial_t \omega + (\omega \cdot \nabla) \vec{u} 
= -\alpha \omega + \nu \Delta \omega + (\nabla \times \vec{f})_z  \ ,
\eeq
where $\vec{u} = (u_x (x,y), u_y (x,y), 0)$ is the velocity field per unit 
mass, $\omega$ the non-vanishing component of its vorticity 
$\nabla \times \vec{u} = (0,0,\omega)$, $\nu$ the kinematic viscosity, 
$\alpha \geqslant 0$ the Ekman friction coefficient  
and $\vec{f}$ a solenoidal body force. The subscripts $x$, $y$ and $z$ denote 
the respective components of a 3d vector field. 

We carry out direct numerical simulations of Eq.~\eqref{eq:momentum}  
on $V=[0,2\pi]^2$ using the standard pseudospectral method \citep{Orszag69} 
for spatial 
discretization in conjunction with full dealiasing by truncation 
following the $2/3$rds rule \citep{Orszag71}. The initial data consist of  
random, Gaussian distributed vorticity fields. 
Owing to the focus on condensate formation and its dependence on large-scale 
friction, the friction coefficient $\alpha$ was small or set to zero in some of the simulations. 
In the latter case the condensate saturates on a viscous time scale \citep{Chan12,Linkmann19b}
with the consequence that the simulations
need to be evolved for a long time in order to obtain statistically 
stationary states. Similarly long transients occur for low values of $\alpha$.
As such, it was necessary to compromise on resolution,  
and the simulations were run using $256^2-512^2$ grid points. 

In order to study transitions to condensate formation, we conduct a parameter 
study for a stochastic, Gaussian distributed and $\delta$-in-time correlated force 
$\vec{f}_s$ that is applied at scales corresponding to a wavenumber interval 
$[k_{\rm min},k_{\rm max}]$, and compare the results with those obtained with a 
forcing that is linear in the velocity field \citep{Linkmann19a,Linkmann19b}, i.e. 
%
\beq
\label{eq:linforce1}
\fvec{f}_l(\vec{k})  = \nu_i k^2 \gamma_k \fvec{u}(\vec{k}) \ ,
\eeq
where $\gamma_k$ is a spherically symmetric Galerkin projector
\beq
\label{eq:linforce2}
\gamma_k =
\begin{cases}
        1 \quad \text{ for } k \in [k_{\rm min},k_{\rm max}] \ , \\
        0 \quad \text{ otherwise} \ .
\end{cases}
\eeq
Here $k = |\vec{k}|$, while $\hat\cdot$ denotes the Fourier transform and
$\nu_i > 0$ an amplification factor, such that the driving occurs through a
linear instability in the wavenumber interval $[k_{\rm min},k_{\rm max}]$.  
For numerical stability an additional dissipation term 
$\tilde \nu \Delta \omega$ is used at small scales, i.e. at $k > k_{\rm min}$, 
where the parameter $\tilde \nu = (\nu_2-\nu) > 0$ mimicks the effect of 
hyperviscosity.	
The
linear forcing is inspired by single-equation models describing dense bacterial
suspensions \citep{Wensink12,Slomka15EPJST,Linkmann19a,Linkmann19b}, where {\em
active turbulence} occurs. The latter is a spatiotemporally chaotic state
characterised by the formation of mesoscale vortices owing to the collective
effects of the microswimmers. These vortices occur in a narrow band of length
scales, and can be described through a linear instability in the wavenumber
interval $[k_{\rm min},k_{\rm max}]$ \citep{Wensink12,Slomka15EPJST,Linkmann19b}. 
For both $\vec{f}_s$ and $\vec{f}_l$, statistically stationary states are eventually reached, 
where the spatiotemporally averaged energy dissipation, $\eps$,
balances the spatiotemporally averaged energy input, $\epsin$,	
\beq
\eps := \langle \eps(t)\rangle_t = \nu \langle \omega^2 \rangle_{V,t} + \alpha \langle |\vec{u}|^2 \rangle_{V,t} = \langle \vec{f} \cdot \vec{u} \rangle_{V,t} = \langle \epsin(t)\rangle_t=: \epsin\ ,
\eeq
with $\langle \cdot \rangle_{V,t} = \langle \langle \cdot \rangle_V \rangle_{t}$ denoting the combined spatial and temporal average.
For Gaussian-distributed and $\delta(t)$-correlated random forcing, 
$\epsin$ is known {\em a priori} \citep{Novikov65}
\beq
{\epsin}_s = \langle \vec{f}_s \cdot \vec{u} \rangle_{V,t} = \frac{F^2}{2} \ ,
\eeq
where $F = \langle|\vec{f}_s|\rangle_{V,t}$.
That is, the energy input is a control parameter rather than 
an observable in simulations using $\vec{f}_s$. 
Details of the simulations are summarised in table \ref{tbl:simulations}. 

For the linear forcing, the energy input is 
\beq
{\epsin}_l(t) = 2 (\nu_i - \nu) \int_{k_{\rm min}}^{k_{\rm max}} dk \ k^2 E(k,t) \ , 
\eeq
where 
\beq
E(k,t) = \frac{1}{2} \int_{|\vec{k}| = k} d\vec{k} \ |\fvec{u}(\vec{k},t)| \ ,  
\eeq
is the energy spectrum.  
Equation \eqref{eq:momentum} with $\vec{f} = \vec{f}_l$ and the aformentioned 
enhanced small-scale damping has been solved 
numerically by \citet{Linkmann19a,Linkmann19b} in the context of transitions to 
large-scale pattern formation in dense suspensions of active matter. 
Here, we compare our simulations listed in table \ref{tbl:simulations} 
against the data of \citet{Linkmann19a,Linkmann19b}, which is 
summarised in table \ref{tbl:simulations-prl}. 
All simulations are evolved for several thousand large-eddy turnover times 
$\tau = L/U$, where $U$ is the root-mean-square velocity and  
$L=2/U^2 \int_0^\infty dk \ E(k)/k$ the integral length scale, with 
$E(k) = \langle E(k,t)\rangle_t$.

\begin{table}
	\begin{tabular}{ccccccccc}
		\hline
		$N$ & $F$& $\alpha $ & Re & Re$_f$ & $U$ & $U_f$ & $L$ & $\tau $  \\
		\hline
		256 & 0.08-0.23 & 0       & 21-11817   & 9.4 - 11.8  & 0.09 - 3.05  & 0.06 - 0.07  & 0.11 - 1.94 & 1.21 - 0.64 \\
		256 & 0.10-0.29 & 0.0005  & 19 - 10083 & 9.4 - 15.0  & 0.09 - 2.67  & 0.06 - 0.09 & 0.11 - 1.89 & 1.24 - 0.71  \\
		256 & 0.10-0.32 & 0.001   & 19 - 8719  & 9.4 - 16.8 & 0.09 - 2.36 & 0.06 - 0.10 & 0.11 - 1.85 & 1.24 - 0.78  \\
		256 & 0.10-0.32 & 0.005   & 17 - 3592  & 9.5 - 18.3 & 0.09 - 1.08 & 0.06 - 0.11 & 0.10 - 1.67 & 1.15 - 1.55  \\
		\hline
		512 & 0.11 & 0 & 29.1& 9.79 & 0.106 & 0.057 & 0.14 & 1.32 \\
		512 & 0.14 & 0 & 2321.5& 10.14 & 0.627 & 0.059 & 1.85 & 2.95  \\
		512 & 0.23 & 0 & 12587.2& 11.45 & 3.242 & 0.067 & 1.94 & 0.60  \\
		\hline
	\end{tabular}
	\caption{Simulation details, where $N$ is the number of grid points in each coordinate, 
	$F$ the magnitude of the force acting in the interval $[k_{\rm min}, k_{\rm max}]$ 
	with $k_{\rm min} = 33$ and $k_{\rm max} = 40$ for all simulations, $\alpha$ the large-scale friction parameter, 
	Re $=UL/\nu$ the Reynolds number with respect to the root-mean-square velocity $U$, 
	the integral length scale $L=2/U^2 \int_0^\infty dk \ E(k)/k$ and the kinematic viscosity $\nu$. 
	The latter was set to $\nu = 0.0005$ for all simulations. The Reynolds number at the driven scales is 
	Re$_f = U_fL_f/\nu$, with 
	$U_f = \left(\int_{k_{\rm min}}^{k_{\rm max}} dk \ E(k)\right)^{1/2}$ denoting the velocity at the driven 
	scales and $L_f = 2\pi/(k_{\rm min}+ k_{\rm max})$. 
	The large-eddy turnover time is denoted by $\tau = L/U$, and $\eta =(\nu^3/\varepsilon)^{1/4}$.
	}
	\label{tbl:simulations}
\end{table}

\begin{table}
\centering
\begin{tabular}{cccccccccc}
\hline
	$N$ & $(\nu_i-\nu)/\nu$ & $\tilde\nu/\nu$ & $k_{\rm min}$ & $k_{\rm max}$ & Re & Re$_f$ & $U$ & $L$ & $\tau$ \\
\hline
	 256 & 0.25 - 7.0 & 10.0 &  33 &  40 & 19 - 13677 & 14 - 21 & 0.29 - 7.77 & 0.07 - 1.92 & 0.24 - 2.29\\
	1024 & 1.0      & 10.0 & 129 & 160 & 45       & 21 & 0.027     & 0.029     & 1.07 \\
	1024 & 2.0      & 10.0 & 129 & 160 & 226      & 20 & 0.041     & 0.094     & 2.29 \\
	1024 & 5.0      & 10.0 & 129 & 160 & 132914   & 15 & 1.17      & 1.93      & 1.65 \\
\hline
\end{tabular}
 \caption{
Parameters used in DNSs of the piecewise constant viscosity model and resulting
observables \citep{Linkmann19a,Linkmann19b}. The number of grid points in each coordinate is denoted by $N$,
the viscosity by $\nu$, 
and $\nu_i$, 
$k_{\rm min}$, $k_{\rm max}$ are the parameters in Eqs.~\eqref{eq:linforce1}-\eqref{eq:linforce2}. 
The Reynolds number
${\rm Re} = UL/\nu$ is based on $\nu$, the root-mean-square velocity $U$
and the integral length scale $L = 2/U^2 \int_0^\infty dk \ E(k)/k$,
with $\nu = 1.1 \times 10^{-3}$  for $N = 256$ and
$\nu = 1.7 \times 10^{-5}$  for $N = 1024$.
The parameter $\tilde \nu = (\nu_2-\nu)$ is an additional dissipation term
that mimicks the effect of hyperviscosity at $k > k_{\rm min}$. The large-scale 
friction parameter $\alpha = 0$ for all simulations.  
Averages in the statistically stationary state are computed from at least
1800 snapshots separated by one large-eddy turnover time $\tau=L/U$.
 }
 \label{tbl:simulations-prl}
\end{table}

\section{Random forcing}
\begin{figure}
	\begin{minipage}[l]{.7\columnwidth}
		\includegraphics[width=1.1\textwidth]{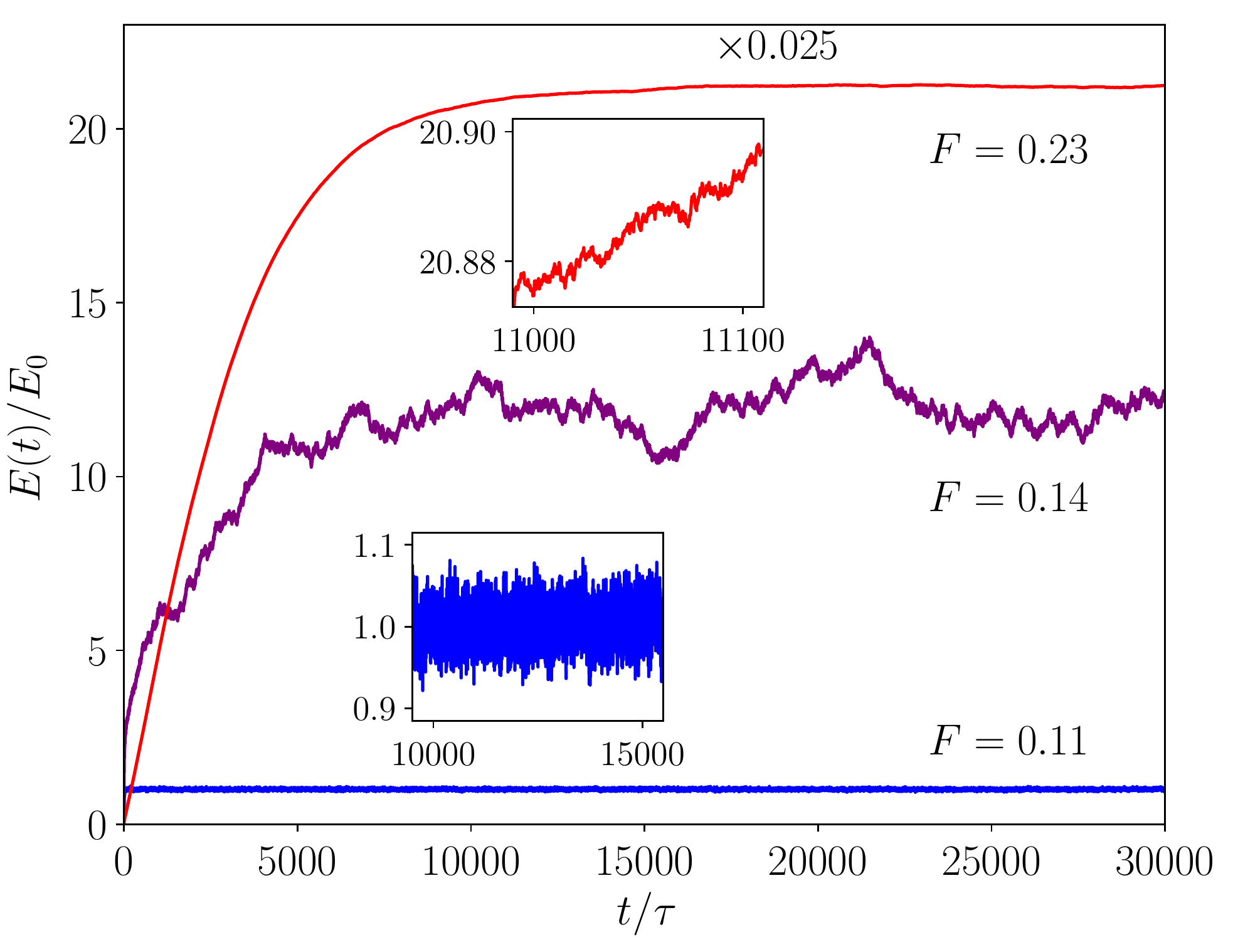} 
	\end{minipage}
	\hspace{3em}
	\begin{minipage}[r]{.2\columnwidth}
		\vspace{-2em}
	\includegraphics[width=.9\textwidth]{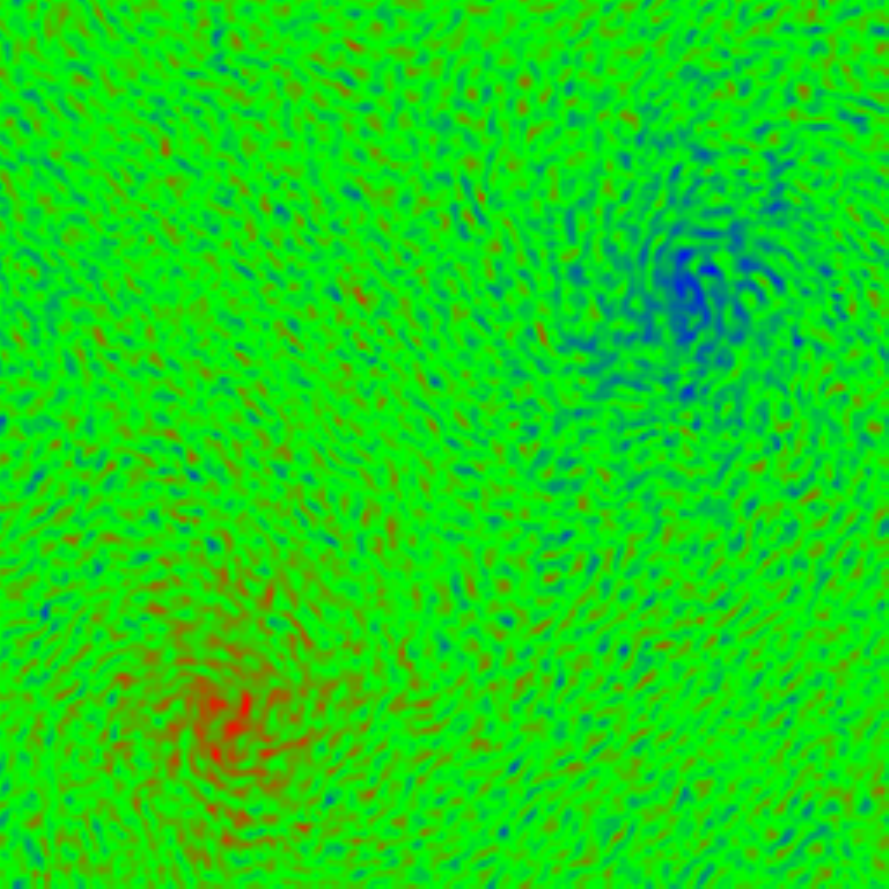} 
		\includegraphics[width=.9\textwidth]{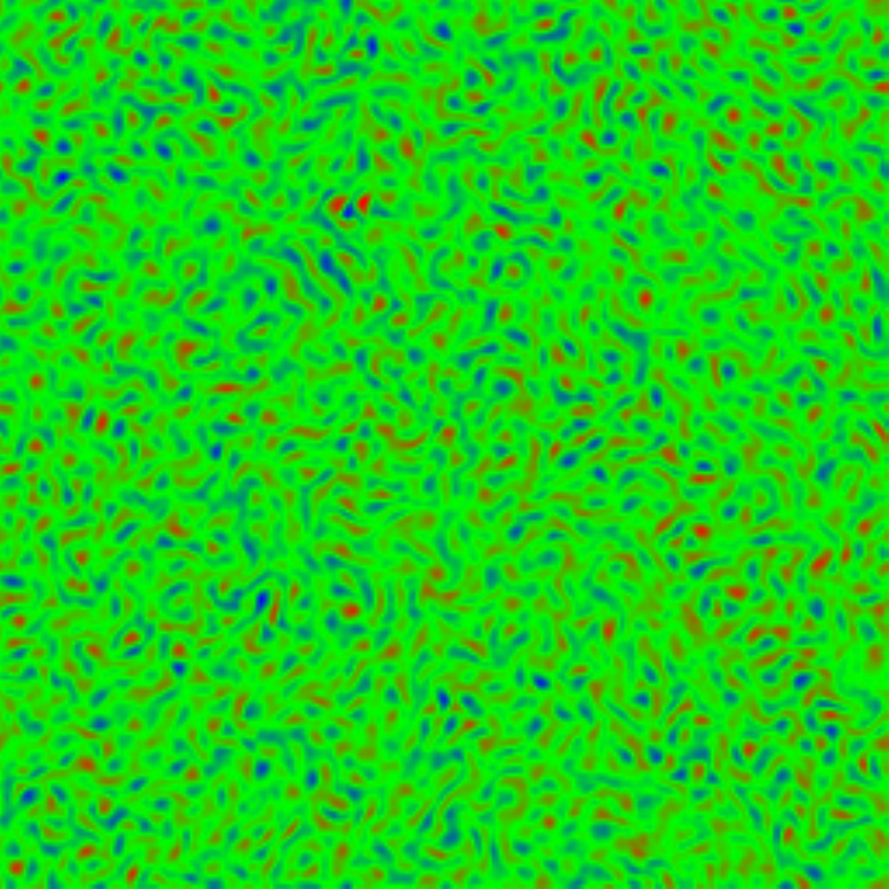} 
		\includegraphics[width=.9\textwidth]{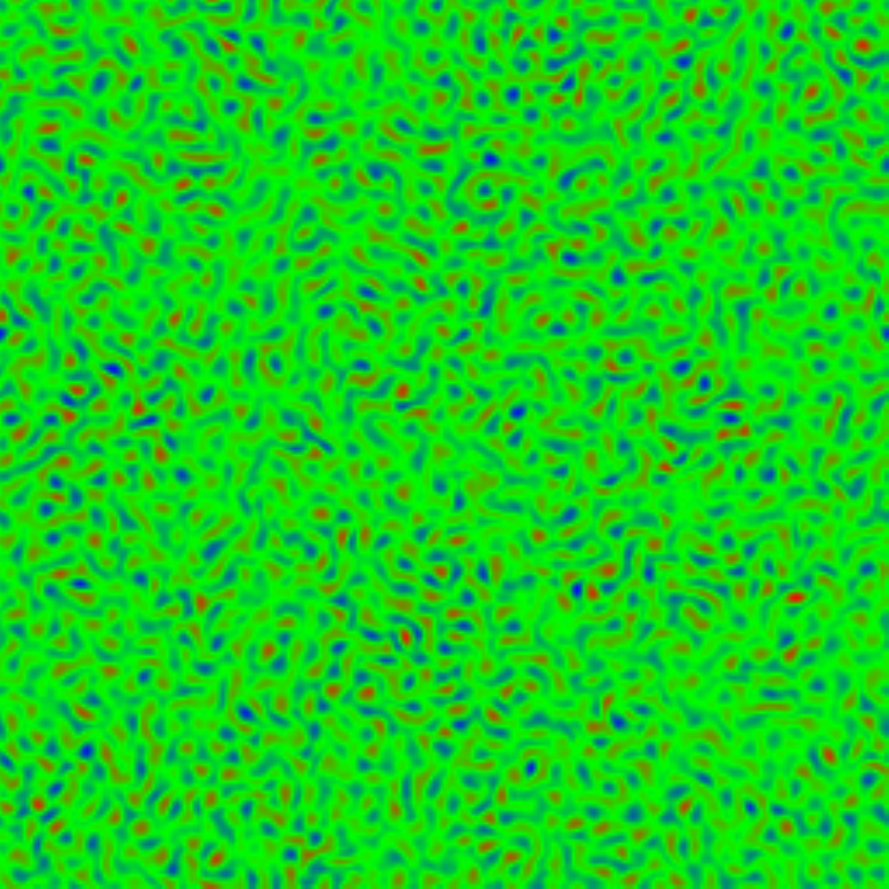} 
	\end{minipage}
	\caption{Left: Time series for $\alpha = 0$ and  
	$F = 0.11$ (blue), $F = 0.14$ (purple) 
	and $F = 0.23$ (red). The data has been normalised with respect to 
	the time averaged energy for $F = 0.11$, $E_0$, and the data for 
	$F = 0.23$ has been further 
	divided by a factor 40 for presentational purposes. Time is given 
	in units of large-eddy turnover time $\tau$.
	Right: Corresponding visualisations of 
	the respective vorticity fields during statistically stationary evolution.} 
	\label{fig:visu}
\end{figure}

Before reporting on the results from the parameter study varying $F$, we briefly discuss
dynamical and statistical properties of the simulated flows using three example cases
with $F = 0.11$, $F = 0.14$ and $F = 0.23$.
Time series of the kinetic energy and visualisations of the vorticity field corresponding to these three cases
are shown in Fig.~\ref{fig:visu}. For $F=0.23$ a condensate 
consisting of two counter-rotating vortices has formed. The remaining cases
do not show large-scale structure formation.

\begin{figure}
	\includegraphics[width=0.49\columnwidth]{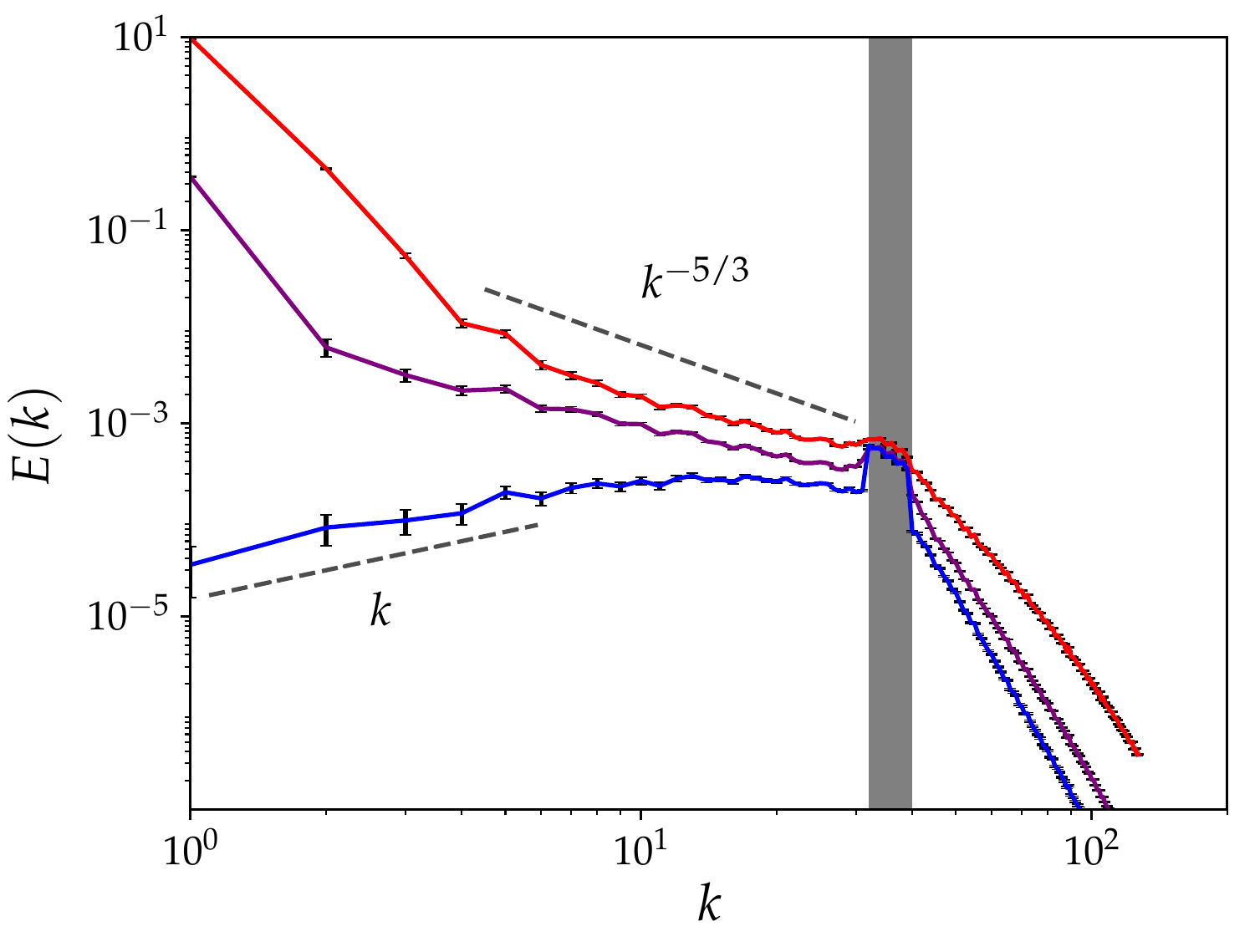} 
	\includegraphics[width=0.49\columnwidth]{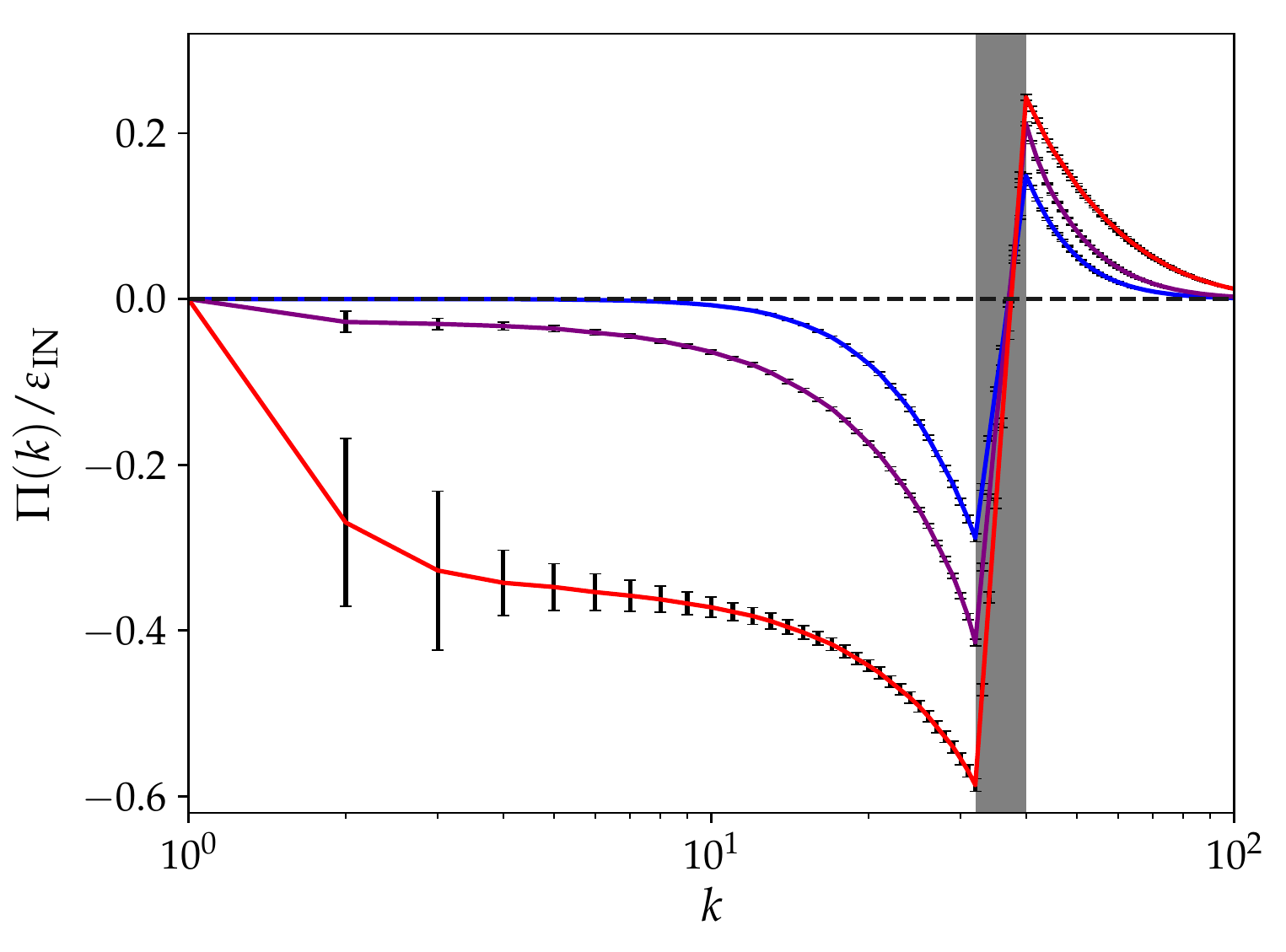} 
	\caption{Energy spectra (left) and normalised fluxes (right) for 
	$\alpha = 0$ and $F = 0.11$ (red), $F = 0.14$ (purple) 
	and $F = 0.23$ (blue) for higher-resolved data in table \ref{tbl:simulations}. 
	The grey-shaded area indicates the 
	driving range. The error bars indicate the standard error calculated from statistically 
	independent samples. 
	}
	\label{fig:rand-espec-flux}
\end{figure}

Figure \ref{fig:rand-espec-flux} presents energy spectra (left) and 
normalised fluxes (right) for $F = 0.11$, $F = 0.14$ and 
$F = 0.23$. A scaling range characterised by a 
scaling exponent of the energy spectrum close to the 
Kolmogorov-value $-5/3$ and a nearly 
wavenumber-independent flux only form at the largest value of $F$. 
For smaller $F$ the flux tends to zero rapidly for $k<k_{\rm min}$, 
hence dissipation cannot be negligible in this wavenumber range. 
In all cases the maximum and minimum values of the normalised flux do not 
add up to unity, which indicates that a substantial amount of energy is
dissipated directly in the driving range. 
Interestingly, for the intermediate $F$, the scaling exponent of 
the energy spectrum is still close but slightly larger than $-5/3$. For the smallest value 
of $F$ the energy spectrum scales linearly with $k$ for $k<7$, 
indicative of energy equipartition among Fourier modes in this wavenumber 
range.
A similar transition in the energy spectra in statistically stationary 2d turbulence 
occurs if the condensate is avoided through a strong drag term \citep{Tsang09}, 
in the sense that the extent of the $-5/3$ scaling range decreases with increasing 
large-scale friction and a power law with positive exponent appears at low wavenumbers. 
However, as a drag term alters the scale-by-scale energy balance, it breaks the 
zero-flux equilibrium condition that underlies linear scaling in 2d,
the low-wavenumber scaling in presence of drag is expected to  
differ from the absolute equilibrium scaling observed here for $\alpha = 0$. 
Indeed, in the former case the spectra are much steeper \citep{Tsang09}.  

Condensates are known to affect inertial-range physics in terms of the
properties of the third-order structure function \citep{Xia08} and the scaling
of the energy spectrum in the inertial range of scales \citep{Chertkov07}.  The
spectral slopes observed here for the random forcing case and in presence of a
condensate are similar to those reported by \citet{Linkmann19b} for the
linearly forced case, hence the details of the small-scale forcing do not
affect the spectral exponent. Deviations of the spectral exponent from 
the Kolmogorov value occur in a variety of turbulent systems. 
For a modified version of the Kuramoto-Sivashinski equation that 
allowed systematic deviations from inertial transfer, 
\citet{Bratanov13} showed by semi-analytical and numerical means 
that nonuniversal power laws arise in spectral intervals where 
the ratio of linear and non-linear time scales is wavenumber-independent. 
As strong condensates result in a significant contribution of linear terms to the 
dynamics, a similar analysis could potentially lead to further insights on the 
nonuniversal scaling exponents in 2d turbulence.

\section{Nonuniversal transitions}
\label{sec:rand-transition}
The transition to condensate formation in 2d turbulence as a function of the energy input 
has so far only been investigated for a single-equation model of active matter 
\citep{Linkmann19a,Linkmann19b}. Here, we now study the transition for a different
kind of forcing and in presence of large-scale friction, as
condensates also occur in presence of drag \citep{Danilov01,Tsang09}. 

\begin{figure}
	\includegraphics[width=0.49\columnwidth]{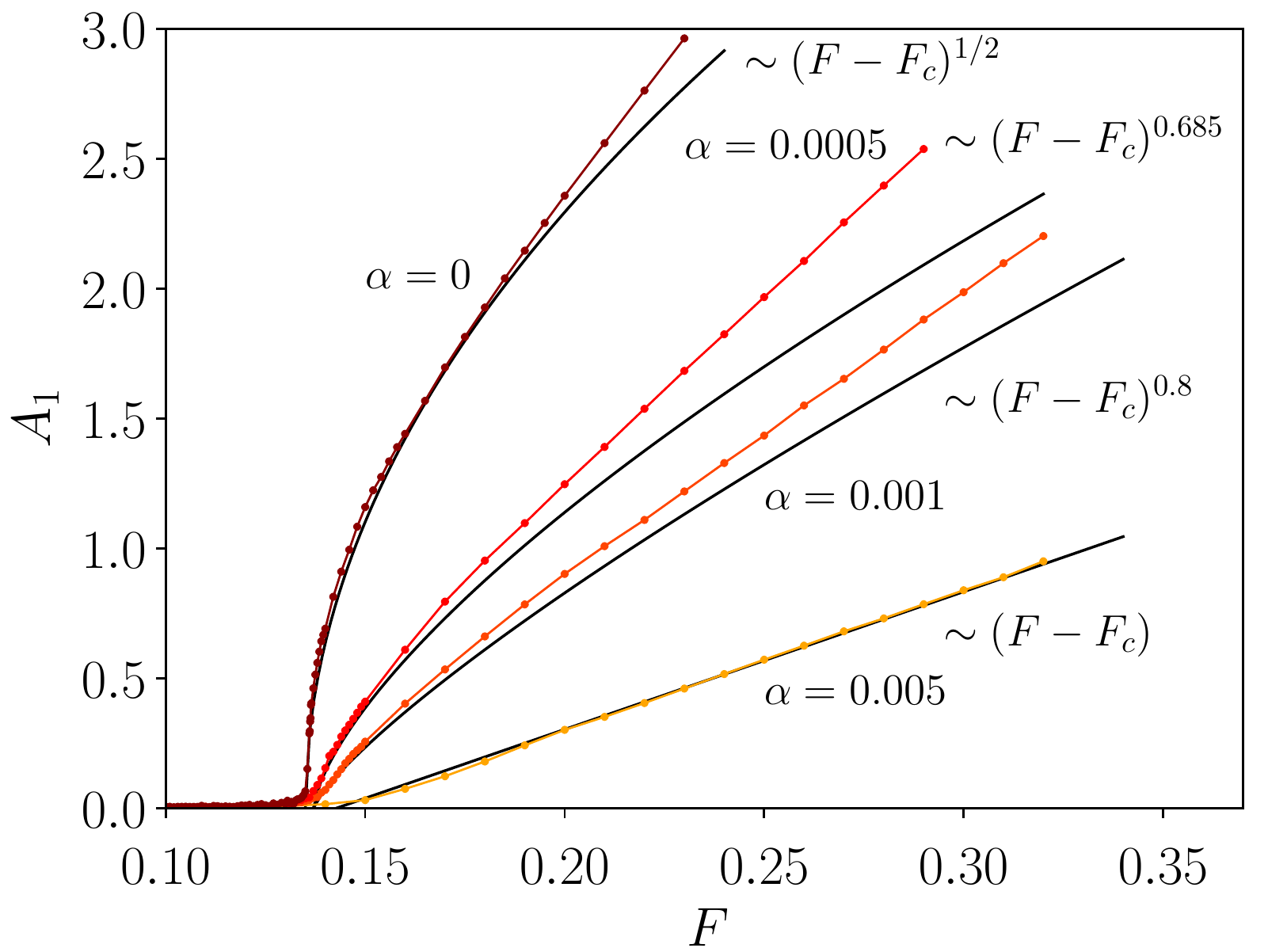} 
	\includegraphics[width=0.49\columnwidth]{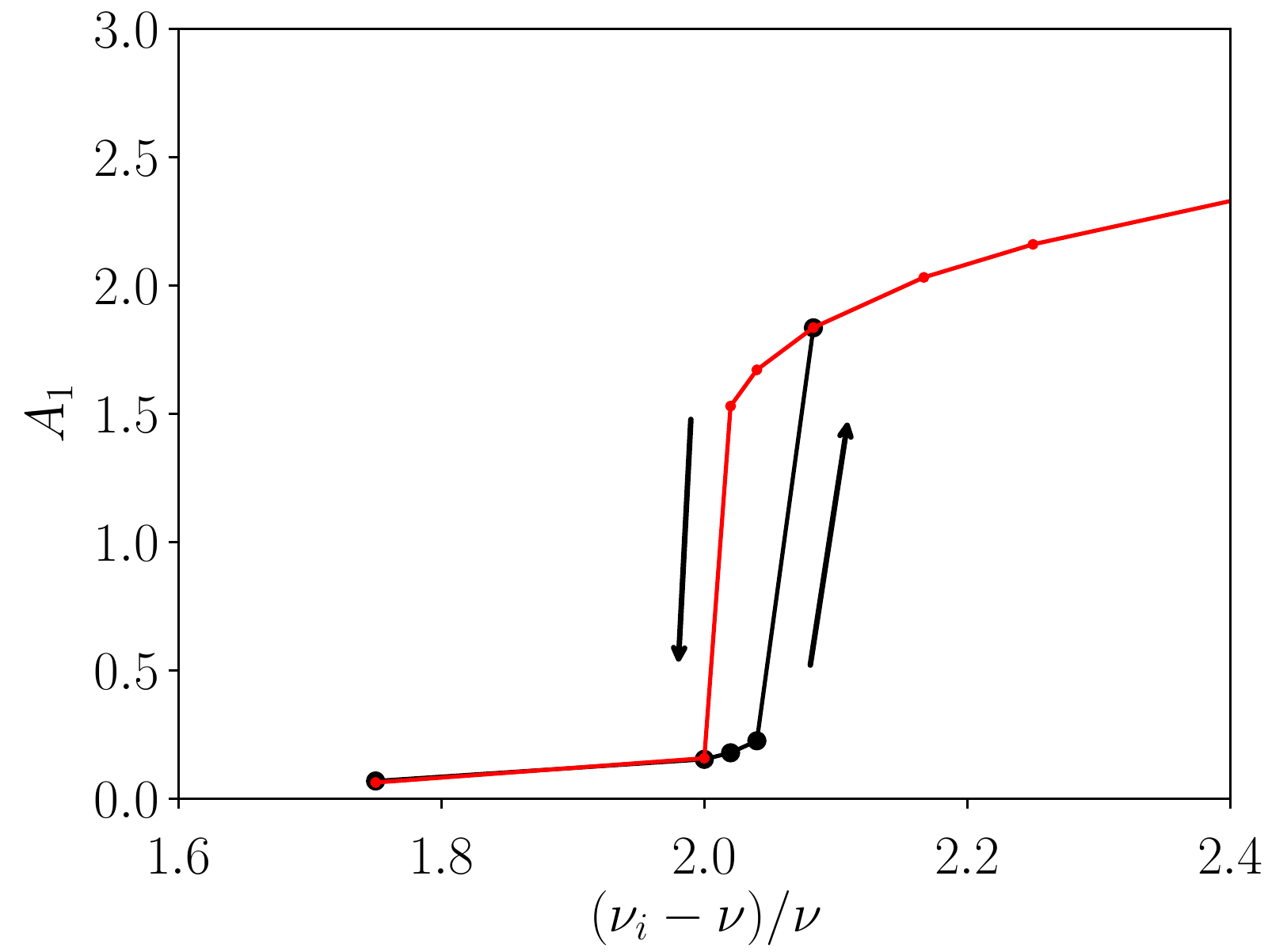} 
	\caption{Amplitude of the Fourier mode at the largest scale, $ A_1 = \sqrt{E(k)_{k=1}}$,  
	as a function of $F$ for random forcing (left) and $\nu_i$ for 
	linear forcing (right).}
	\label{fig:rand-transition}
\end{figure}

The left panel of Fig.~\ref{fig:rand-transition} presents the amplitude of the 
lowest Fourier mode, i.e. the square root of the average energy at the
largest scale, $A_1=\sqrt{E(k)_{k=1}}$, as a function of $F$ from the parameter
study for the random, Gaussian distributed and $\delta(t)$-correlated forcing.
Three main observations can be made from the left panel of the figure. First, there is a clear 
transition point, below which $A_1 \simeq 0$ and above which 
$A_1$ grows with increasing $F$, indicating the formation of a condensate. 
Second, the data appears to be continuous at the critical point $F_c$ with a possibly 
discontinuous first derivative. The critical point is approached from above by a power law
\beq
A_1 \sim (F-F_c)^{\gamma} \ ,  
\eeq
where $F_c = F_c(\alpha)$ and $\gamma = \gamma(\alpha)$ depend on the value of the 
large-scale friction coefficient. 
For $\alpha = 0$ the functional form $A_1(F)$ corresponds to the upper branch of the 
normal form of a supercritical pitchfork bifurcation, that is $\gamma = 1/2$, even though the 
stochastic nature of the system makes comparisons to concepts from dynamical systems 
theory difficult. 
A least-squares fit of $A_1$ against $F$ places the critical 
point at $F_c = 0.135$.  
For $\alpha = 0.0005$ we have $F_c = 0.137$ and $\gamma = 0.685$, 
$\alpha = 0.001$ results in $F_c = 0.137$ and $\gamma = 0.8$ and
$\alpha = 0.005$ corresponds to $F_c = 0.143$ and $\gamma = 1$.
That is, while the approach to the critical point is strongly dependent
on the level of large-scale friction, the location of the critical point varies little. 
Third, for $F \gg F_c$ the amplitude $A_1$ grows linearly with $F$ in all cases. 
Equivalently, 
$E(k)_{k=1}$ grows linearly with $\epsin$, 
which is expected for a sizeable condensate as most of the dissipation should then take 
place at the largest scales
\beq
\epsin = \eps = 2 \nu \int_0^\infty dk \ k^2 E(k) + \alpha \int_0^\infty dk \ E(k) \approx (2 \nu + \alpha) E(k)_{|k=1}\delta k \ ,
\eeq
where $\delta k$ is the grid spacing in Fourier space. The same argument suggests that one can expect 
the linear scaling $A_1 \sim (F-F_c)$ observed for $\alpha = 0.001$ asymptotically for strong friction. 

The type of transition is very different if the driving occurs through a small-scale 
linear instability. The right panel of Fig.~\ref{fig:rand-transition} shows $A_1$ from the data
of \citet{Linkmann19a,Linkmann19b} as a function of $\nu_i$ for the linear forcing 
specified in Eq.~\eqref{eq:linforce1}. In contrast to the randomly forced case, the 
transition is now subcritical \citep{Linkmann19a,Linkmann19b} as evidenced by a discontinuity 
in the data and the clearly visible hysteresis loop. 
As the hysteresis loop is small, one 
may expect that the transitions happen at comparable values of a forcing-scale 
Reynolds number $\Rey_f = U_f L_f/\nu$, where $L_f$ is a length scale that corresponds to the 
middle wavenumber in the driven range, $L_f = 2\pi/(k_{\rm min} + k_{\rm max})$, 
and $U_f$ is the rms velocity in that range of scales
\beq
U_f = \left( \int_{k_{\rm min}}^{k_{\rm min}} dk \ E(k) \right)^{1/2} \ .
\eeq
This is indeed the case, the transition occurs at $\Rey_f \approx 15-20$ in the subcritical case 
\citep{Linkmann19a} and at $\Rey_f \approx 10$ in the supercritical case studied here. 

\section{Conclusions}
\label{sec:conclusions}
We here study the formation of the condensate as a function of the type and
amplitude of the forcing. Direct numerical simulations show that the condensate
does not appear gradually but in a phase transition. For prescribed energy
dissipation the transition is second order, 
and both the critical point and the critical exponent depend on the value of 
the large-scale friction coefficient. 
In this context, we point 
out that $\varepsilon$ does not depend on $\alpha$ for the random, 
$\delta(t)$-correlated forcing used here, as is the case for time-independent
forcing such as Kolmogorov flow \citep{Tsang09}. However, a series of test simulations
using time-independent forcing led to similar results (Musacchio \& Boffetta, private communication).  
When the forcing is due to a small-scale instability as inspired
by continuum models of active matter, the
transition is first order \citep{Linkmann19a,Linkmann19b}.  
The phase transitions separate
two markedly different types of 2d dynamics: in turbulence with a condensate,
energy input is predominantly balanced by dissipation in the condensate and
intermediate scales follow an inertial cascade; without a condensate
dissipation is spread over the intermediate scales and the properties of the
energy transfer are noticeably different and nonuniversal.

In summary, the transition to condensate formation in 2d turbulence is nonuniversal 
in the sense that (i) the type of transition depends on the type of forcing, 
and (ii) the details of the transition for a given type of forcing depend on 
other system parameters such as large-scale friction. 
The presence of these nonuniversalities naturally motivate
questions concerning their origin. Results from rapidly rotating 
RBC \citep{Favier19} suggest that the hysteretic transition in the linearly forced
case may be related to persistent phase correlations between the driven scales and 
the condensate. Random forcing precludes such a scenario. Further questions concern
the theoretical predictions on the dependence of the critical exponent $\gamma$ on the 
level of large-scale friction. The value $\gamma = 1$ is plausible 
for strong linear damping by the same argument that predicted a linear dependence
of the energy in the condensate on the energy input.

\section*{Acknowledgements}
Bruno Eckhardt sadly passed away before the manuscript was written. We hope to have summarised the 
collaborative work according to his standards, and any shortcomings should be attributed to ML.
We will remember him as an outstanding scientist, thoughtful supervisor and inspiring role model.
ML thanks Guido Boffetta and Stefano Musacchio for helpful discussions.

\bibliographystyle{jfm}
\bibliography{references}

\begin{thebibliography}{43}
\expandafter\ifx\csname natexlab\endcsname\relax\def\natexlab#1{#1}\fi
\def\au#1{#1} \def\ed#1{#1} \def\yr#1{#1}\def\at#1{#1}\def\jt#1{\textit{#1}}
  \def\bt#1{#1}\def\bvol#1{\textbf{#1}} \def\vol#1{#1} \def\pg#1{#1}
  \def\publ#1{#1}\def\arxiv#1{#1}\def\org#1{#1}\def\st#1{\textit{#1}}

\bibitem[Alexakis(2015)]{Alexakis15}
{\sc \au{Alexakis, A.}} \yr{2015}  \at{{Rotating Taylor-Green flow}}.  \jt{J.
  Fluid Mech.}  \bvol{769},  \pg{46--78}.

\bibitem[Alexakis \& Biferale(2018)]{Alexakis18}
{\sc \au{Alexakis, A.} \& \au{Biferale, L.}} \yr{2018}  \at{{Cascades and
  transitions in turbulent flows}}.  \jt{Phys. Reports}  \bvol{767-769},
  \pg{1--101}.

\bibitem[Biferale {\em et~al.\/}(2012)Biferale, Musacchio \&
  Toschi]{Biferale12}
{\sc \au{Biferale, L.}, \au{Musacchio, S.} \& \au{Toschi, F.}} \yr{2012}
  \at{{Inverse energy cascade in three-dimensional isotropic turbulence}}.
  \jt{Phys. Rev. Lett.}  \bvol{108},  \pg{164501}.

\bibitem[Boffetta \& Ecke(2014)]{Boffetta14ARFM}
{\sc \au{Boffetta, G.} \& \au{Ecke, R.~E.}} \yr{2014}  \at{{Two-Dimensional
  Turbulence}}.  \jt{Annu. Rev. Fluid Mech.}  \bvol{44},  \pg{427--451}.

\bibitem[Bouchet \& Simonnet(2009)]{Bouchet09}
{\sc \au{Bouchet, F.} \& \au{Simonnet, E.}} \yr{2009}  \at{{Random Changes of
  Flow Topology in Two-Dimensional and Geophysical Turbulence}}.  \jt{Phys.
  Rev. Lett.}  \bvol{102},  \pg{094504}.

\bibitem[Bratanov {\em et~al.\/}(2013)Bratanov, Jenko, Hatch \&
  Wilczek]{Bratanov13}
{\sc \au{Bratanov, V.}, \au{Jenko, F.}, \au{Hatch, D.~R.} \& \au{Wilczek, M.}}
  \yr{2013}  \at{Nonuniversal power-law spectra in turbulent systems}.
  \jt{Phys. Rev. Lett.}  \bvol{111},  \pg{075001}.

\bibitem[Celani {\em et~al.\/}(2010)Celani, Musacchio \& Vincenzi]{Celani10}
{\sc \au{Celani, A.}, \au{Musacchio, S.} \& \au{Vincenzi, D.}} \yr{2010}
  \at{{Turbulence in More Than Two and Less Than Three Dimensions}}.  \jt{Phys.
  Rev. Lett.}  \bvol{104},  \pg{184506}.

\bibitem[Chan {\em et~al.\/}(2012)Chan, Mitra \& Brandenburg]{Chan12}
{\sc \au{Chan, C.}, \au{Mitra, D.} \& \au{Brandenburg, A.}} \yr{2012}
  \at{{Dynamics of saturated energy condensation in two-dimensional
  turbulence}}.  \jt{Phys. Rev. E}  \bvol{85},  \pg{036315}.

\bibitem[Chertkov {\em et~al.\/}(2007)Chertkov, Connaughton, Kolokolov \&
  Lebedev]{Chertkov07}
{\sc \au{Chertkov, M.}, \au{Connaughton, C.}, \au{Kolokolov, I.} \&
  \au{Lebedev, V.}} \yr{2007}  \at{{Dynamics of Energy Condensation in
  Two-Dimensional Turbulence}}.  \jt{Phys. Rev. Lett.}  \bvol{99},
  \pg{084501}.

\bibitem[Danilov \& Gurarie(2001)]{Danilov01}
{\sc \au{Danilov, S.} \& \au{Gurarie, D.}} \yr{2001}  \at{{Forced
  two-dimensional turbulence in spectral and physical space}}.  \jt{Phys. Rev.
  E}  \bvol{63},  \pg{061208}.

\bibitem[Deusebio {\em et~al.\/}(2014)Deusebio, Boffetta, Lindborg \&
  Musacchio]{Deusebio14}
{\sc \au{Deusebio, E.}, \au{Boffetta, G.}, \au{Lindborg, E.} \& \au{Musacchio,
  S.}} \yr{2014}  \at{{Dimensional transition in rotating turbulence}}.
  \jt{Phys. Rev. E}  \bvol{90},  \pg{023005}.

\bibitem[Dombrowski {\em et~al.\/}(2004)Dombrowski, Cisneros, Chatkaew,
  Goldstein \& Kessler]{Dombrowski04}
{\sc \au{Dombrowski, C.}, \au{Cisneros, L.}, \au{Chatkaew, S.}, \au{Goldstein,
  R.~E.} \& \au{Kessler, J.~O.}} \yr{2004}  \at{{Self-Concentration and
  Large-Scale Coherence in Bacterial Dynamics}}.  \jt{Phys. Rev. Lett.}
  \bvol{93},  \pg{098103}.

\bibitem[Dunkel {\em et~al.\/}(2013)Dunkel, Heidenreich, Drescher, Wensink,
  B\"ar \& Goldstein]{Dunkel13PRL}
{\sc \au{Dunkel, J.}, \au{Heidenreich, S.}, \au{Drescher, K.}, \au{Wensink,
  H.~H.}, \au{B\"ar, M.} \& \au{Goldstein, R.~E.}} \yr{2013}  \at{Fluid
  dynamics of bacterial turbulence}.  \jt{Phys. Rev. Lett.}  \bvol{110},
  \pg{228102}.

\bibitem[Favier {\em et~al.\/}(2019)Favier, Guervilly \& Knobloch]{Favier19}
{\sc \au{Favier, B.}, \au{Guervilly, C.} \& \au{Knobloch, E.}} \yr{2019}
  \at{{Subcritical turbulent condensate in rapidly rotating Rayleigh-B\'enard
  convection}}.  \jt{J. Fluid Mech.}  \bvol{864},  \pg{R1}.

\bibitem[Frisch {\em et~al.\/}(1975)Frisch, Pouquet, L\'{e}orat \&
  Mazure]{Frisch75}
{\sc \au{Frisch, U.}, \au{Pouquet, A.}, \au{L\'{e}orat, J.} \& \au{Mazure, A.}}
  \yr{1975}  \at{{Possibility of an inverse cascade of magnetic helicity in
  magnetohydrodynamic turbulence}}.  \jt{J. Fluid Mech.}  \bvol{68},
  \pg{769--778}.

\bibitem[Frishman {\em et~al.\/}(2017)Frishman, Laurie \&
  Falkovich]{Frishman2017}
{\sc \au{Frishman, A.}, \au{Laurie, J.} \& \au{Falkovich, G.}} \yr{2017}
  \at{{Jets or vortices - what flows are generated by an inverse turbulent
  cascade?}}  \jt{Phys. Rev. Fluids}  \bvol{2},  \pg{032602}.

\bibitem[Gachelin {\em et~al.\/}(2014)Gachelin, Rousselet, Lindner \&
  Clement]{Gachelin14}
{\sc \au{Gachelin, J.}, \au{Rousselet, A.}, \au{Lindner, A.} \& \au{Clement,
  E.}} \yr{2014}  \at{Collective motion in an active suspension of {\it
  escherichia coli} bacteria}.  \jt{New Journal of Physics}  \bvol{16},
  \pg{025003}.

\bibitem[Gallet(2015)]{Gallet15b}
{\sc \au{Gallet, B.}} \yr{2015}  \at{{Exact two-dimensionalization of rapidly
  rotating large Reynolds-number flows}}.  \jt{J. Fluid Mech.}  \bvol{783},
  \pg{412--447}.

\bibitem[Gallet \& Doering(2015)]{Gallet15a}
{\sc \au{Gallet, B.} \& \au{Doering, C.~R.}} \yr{2015}  \at{{Exact
  two-dimensionalization of low-magnetic-Reynolds-number flows subject to a
  strong magnetic field}}.  \jt{J. Fluid Mech.}  \bvol{773},  \pg{154--177}.

\bibitem[Hossain {\em et~al.\/}(1983)Hossain, Matthaeus \&
  Montgomery]{Hossain83}
{\sc \au{Hossain, M.}, \au{Matthaeus, W.~H.} \& \au{Montgomery, D.}} \yr{1983}
  \at{Long-time states of inverse cascades in the presence of a maximum length
  scale}.  \jt{J. Plasma Physics}  \bvol{30},  \pg{479–493}.

\bibitem[van Kan \& Alexakis(2019)]{vanKan19}
{\sc \au{van Kan, A.} \& \au{Alexakis, A.}} \yr{2019}  \at{{Condensates in thin
  layer turbulence}}.  \jt{J. Fluid Mech.}  \bvol{864},  \pg{490--518}.

\bibitem[Kraichnan(1967)]{Kraichnan67}
{\sc \au{Kraichnan, R.~H.}} \yr{1967}  \at{Inertial ranges in two‐dimensional
  turbulence}.  \jt{Phys. Fluids}  \bvol{10},  \pg{1417--1423}.

\bibitem[Linkmann {\em et~al.\/}(2019{\natexlab{{\em a\/}}})Linkmann, Boffetta,
  Marchetti \& Eckhardt]{Linkmann19a}
{\sc \au{Linkmann, M.}, \au{Boffetta, G.}, \au{Marchetti, M.~C.} \&
  \au{Eckhardt, B.}} \yr{2019{\natexlab{{\em a\/}}}}  \at{{Phase Transition to
  Large Scale Coherent Structures in Two-Dimensional Active Matter
  Turbulence}}.  \jt{Phys. Rev. Lett.}  \bvol{122},  \pg{214503}.

\bibitem[Linkmann {\em et~al.\/}(2019{\natexlab{{\em b\/}}})Linkmann,
  Marchetti, Boffetta \& Eckhardt]{Linkmann19b}
{\sc \au{Linkmann, M.}, \au{Marchetti, M.~C.}, \au{Boffetta, G.} \&
  \au{Eckhardt, B.}} \yr{2019{\natexlab{{\em b\/}}}}  \at{{Condensate formation
  and multiscale dynamics in two-dimensional active suspensions}}.
  \jt{arXiv:1905.06267} .

\bibitem[Marino {\em et~al.\/}(2013)Marino, Mininni, Rosenberg \&
  Pouquet]{Marino13}
{\sc \au{Marino, R.}, \au{Mininni, P.~D.}, \au{Rosenberg, D.} \& \au{Pouquet,
  A.}} \yr{2013}  \at{{Inverse cascades in rotating stratified turbulence: fast
  growth of large scales}}.  \jt{Europhys. Lett.}  \bvol{102},  \pg{44006}.

\bibitem[Musacchio \& Boffetta(2017)]{Musacchio17}
{\sc \au{Musacchio, S.} \& \au{Boffetta, G.}} \yr{2017}  \at{{Split energy
  cascade in turbulent thin fluid layers}}.  \jt{Phys. Fluids}  \bvol{29},
  \pg{111106}.

\bibitem[Novikov(1965)]{Novikov65}
{\sc \au{Novikov, E.~A.}} \yr{1965}  \at{{Functionals and the random-force
  method in turbulence theory}}.  \jt{Soviet Physics JETP}  \bvol{20},
  \pg{1290--1294}.

\bibitem[Orszag(1969)]{Orszag69}
{\sc \au{Orszag, S.~A.}} \yr{1969}  \at{Numerical methods for the simulation of
  turbulence}.  \jt{Phys. Fluids}  \bvol{12},  \pg{II--250--II--257}.

\bibitem[Orszag(1971)]{Orszag71}
{\sc \au{Orszag, S.~A.}} \yr{1971}  \at{{On the Elimination of Aliasing in
  Finite-Difference Schemes by Filtering High-Wavenumber Components}}.  \jt{J.
  Atmos. Sci.}  \bvol{28},  \pg{1074}.

\bibitem[Pouquet {\em et~al.\/}(1976)Pouquet, Frisch \& L{\'e}orat]{Pouquet76}
{\sc \au{Pouquet, A.}, \au{Frisch, U.} \& \au{L{\'e}orat, J.}} \yr{1976}
  \at{{Strong {M}{H}{D} helical turbulence and the nonlinear dynamo effect}}.
  \jt{J. Fluid Mech.}  \bvol{77},  \pg{321--354}.

\bibitem[Rubio {\em et~al.\/}(2014)Rubio, Julien, Knobloch \& Weiss]{Rubio14}
{\sc \au{Rubio, A.~M.}, \au{Julien, K.}, \au{Knobloch, E.} \& \au{Weiss,
  J.~B.}} \yr{2014}  \at{{Upscale Energy Transfer in Three-Dimensional Rapidly
  Rotating Turbulent Convection}}.  \jt{Phys. Rev. Lett.}  \bvol{112},
  \pg{144501}.

\bibitem[Seshasayanan \& Alexakis(2018)]{Seshasayanan18}
{\sc \au{Seshasayanan, K.} \& \au{Alexakis, A.}} \yr{2018}  \at{{Condensates in
  rotating turbulent flows}}.  \jt{J. Fluid Mech.}  \bvol{841},  \pg{434--462}.

\bibitem[S\l{}omka \& Dunkel(2015)]{Slomka15EPJST}
{\sc \au{S\l{}omka, J.} \& \au{Dunkel, J.}} \yr{2015}  \at{{Generalized
  Navier-Stokes equations for active suspensions}}.  \jt{Eur. Phys. J. Spec.
  Top.}  \bvol{224},  \pg{1349}.

\bibitem[Smith \& Yakhot(1993)]{Smith93}
{\sc \au{Smith, L.~M.} \& \au{Yakhot, V.}} \yr{1993}  \at{{Bose condensation
  and small-scale structure generation in a random force driven 2D
  turbulence}}.  \jt{Phys. Rev. Lett.}  \bvol{71},  \pg{352--355}.

\bibitem[Sommeria(1993)]{Sommeria86}
{\sc \au{Sommeria, J.}} \yr{1993}  \at{{Experimental study of the
  two-dimensional inverse energy cascade in a square box}}.  \jt{J. Fluid
  Mech.}  \bvol{170},  \pg{139--168}.

\bibitem[Sozza {\em et~al.\/}(2015)Sozza, Boffetta, Muratore-Ginanneschi \&
  Musacchio]{Sozza15}
{\sc \au{Sozza, A.}, \au{Boffetta, G.}, \au{Muratore-Ginanneschi, P.} \&
  \au{Musacchio, S.}} \yr{2015}  \at{{Dimensional transition of energy cascades
  in stably stratified forced thin fluid layers}}.  \jt{Phys. Fluids}
  \bvol{27},  \pg{035112}.

\bibitem[Tsang \& Young(2009)]{Tsang09}
{\sc \au{Tsang, Y.-K.} \& \au{Young, W.~R.}} \yr{2009}  \at{{Forced-dissipative
  two-dimensional turbulence: A scaling regime controlled by drag}}.  \jt{Phys.
  Rev. E}  \bvol{79},  \pg{045308(R)}.

\bibitem[Vallis(2006)]{Vallis06}
{\sc \au{Vallis, G.~K.}} \yr{2006} {\em Atmospheric and Oceanic Fluid
  Dynamics\/}.  \publ{Cambridge: Cambridge University Press}.

\bibitem[Waleffe(1993)]{Waleffe93}
{\sc \au{Waleffe, F.}} \yr{1993}  \at{Inertial transfers in the helical
  decomposition}.  \jt{Phys. Fluids A}  \bvol{5},  \pg{677--685}.

\bibitem[Wensink {\em et~al.\/}(2012)Wensink, Dunkel, Heidenreich, Drescher,
  Goldstein, L{\"o}wen \& Yeomans]{Wensink12}
{\sc \au{Wensink, H.~H.}, \au{Dunkel, J.}, \au{Heidenreich, S.}, \au{Drescher,
  K.}, \au{Goldstein, R.~E.}, \au{L{\"o}wen, H.} \& \au{Yeomans, J.~M.}}
  \yr{2012}  \at{Meso-scale turbulence in living fluids}.  \jt{Proc. Natl.
  Acad. Sci.}  \bvol{109},  \pg{14308--14313}.

\bibitem[Xia {\em et~al.\/}(2008)Xia, Punzmann, Falkovich \& Shats]{Xia08}
{\sc \au{Xia, H.}, \au{Punzmann, H.}, \au{Falkovich, G.} \& \au{Shats, M.}}
  \yr{2008}  \at{Turbulence-condensate interaction in two dimensions}.
  \jt{Phys. Rev. Lett.}  \bvol{101},  \pg{194504}.

\bibitem[Xia {\em et~al.\/}(2009)Xia, Shats \& Falkovich]{Xia09}
{\sc \au{Xia, H.}, \au{Shats, M.} \& \au{Falkovich, G.}} \yr{2009}
  \at{Spectrally condensed turbulence in thin layers}.  \jt{Physics of Fluids}
  \bvol{21}~(12),  \pg{125101}.

\bibitem[Yokoyama \& Takaoka(2017)]{Yokoyama17}
{\sc \au{Yokoyama, N.} \& \au{Takaoka, M.}} \yr{2017}  \at{{Hysteretic
  transitions between quasi-two-dimensional flow and three-dimensional flow in
  forced rotating turbulence}}.  \jt{Phys. Rev. Fluids}  \bvol{2},
  \pg{092602(R)}.

\end{thebibliography}

\end{document}